\documentclass{aastex63}
\usepackage{savesym}
\savesymbol{tablenum}
\usepackage{siunitx}
\restoresymbol{SIX}{tablenum}

\newcommand{\mvir}{M_{500c}}
\newcommand{\Yvir}{Y_{500c}}
\newcommand{\snrlimit}{5}
\newcommand{\msol}{M_\odot}

\accepted{\today}

\submitjournal{ApJL}

\graphicspath{{./}{figures/}}

\begin{document}

\title{Searching for High-Energy Neutrino Emission from Galaxy Clusters with IceCube}



\affiliation{III. Physikalisches Institut, RWTH Aachen University, D-52056 Aachen, Germany}
\affiliation{Department of Physics, University of Adelaide, Adelaide, 5005, Australia}
\affiliation{Dept. of Physics and Astronomy, University of Alaska Anchorage, 3211 Providence Dr., Anchorage, AK 99508, USA}
\affiliation{Dept. of Physics, University of Texas at Arlington, 502 Yates St., Science Hall Rm 108, Box 19059, Arlington, TX 76019, USA}
\affiliation{CTSPS, Clark-Atlanta University, Atlanta, GA 30314, USA}
\affiliation{School of Physics and Center for Relativistic Astrophysics, Georgia Institute of Technology, Atlanta, GA 30332, USA}
\affiliation{Dept. of Physics, Southern University, Baton Rouge, LA 70813, USA}
\affiliation{Dept. of Physics, University of California, Berkeley, CA 94720, USA}
\affiliation{Lawrence Berkeley National Laboratory, Berkeley, CA 94720, USA}
\affiliation{Institut f{\"u}r Physik, Humboldt-Universit{\"a}t zu Berlin, D-12489 Berlin, Germany}
\affiliation{Fakult{\"a}t f{\"u}r Physik {\&} Astronomie, Ruhr-Universit{\"a}t Bochum, D-44780 Bochum, Germany}
\affiliation{Universit{\'e} Libre de Bruxelles, Science Faculty CP230, B-1050 Brussels, Belgium}
\affiliation{Vrije Universiteit Brussel (VUB), Dienst ELEM, B-1050 Brussels, Belgium}
\affiliation{Department of Physics and Laboratory for Particle Physics and Cosmology, Harvard University, Cambridge, MA 02138, USA}
\affiliation{Dept. of Physics, Massachusetts Institute of Technology, Cambridge, MA 02139, USA}
\affiliation{Dept. of Physics and The International Center for Hadron Astrophysics, Chiba University, Chiba 263-8522, Japan}
\affiliation{Department of Physics, Loyola University Chicago, Chicago, IL 60660, USA}
\affiliation{Dept. of Physics and Astronomy, University of Canterbury, Private Bag 4800, Christchurch, New Zealand}
\affiliation{Dept. of Physics, University of Maryland, College Park, MD 20742, USA}
\affiliation{Dept. of Astronomy, Ohio State University, Columbus, OH 43210, USA}
\affiliation{Dept. of Physics and Center for Cosmology and Astro-Particle Physics, Ohio State University, Columbus, OH 43210, USA}
\affiliation{Niels Bohr Institute, University of Copenhagen, DK-2100 Copenhagen, Denmark}
\affiliation{Dept. of Physics, TU Dortmund University, D-44221 Dortmund, Germany}
\affiliation{Dept. of Physics and Astronomy, Michigan State University, East Lansing, MI 48824, USA}
\affiliation{Dept. of Physics, University of Alberta, Edmonton, Alberta, Canada T6G 2E1}
\affiliation{Erlangen Centre for Astroparticle Physics, Friedrich-Alexander-Universit{\"a}t Erlangen-N{\"u}rnberg, D-91058 Erlangen, Germany}
\affiliation{Physik-department, Technische Universit{\"a}t M{\"u}nchen, D-85748 Garching, Germany}
\affiliation{D{\'e}partement de physique nucl{\'e}aire et corpusculaire, Universit{\'e} de Gen{\`e}ve, CH-1211 Gen{\`e}ve, Switzerland}
\affiliation{Dept. of Physics and Astronomy, University of Gent, B-9000 Gent, Belgium}
\affiliation{Dept. of Physics and Astronomy, University of California, Irvine, CA 92697, USA}
\affiliation{Karlsruhe Institute of Technology, Institute for Astroparticle Physics, D-76021 Karlsruhe, Germany }
\affiliation{Karlsruhe Institute of Technology, Institute of Experimental Particle Physics, D-76021 Karlsruhe, Germany }
\affiliation{Dept. of Physics, Engineering Physics, and Astronomy, Queen's University, Kingston, ON K7L 3N6, Canada}
\affiliation{Dept. of Physics and Astronomy, University of Kansas, Lawrence, KS 66045, USA}
\affiliation{Department of Physics and Astronomy, UCLA, Los Angeles, CA 90095, USA}
\affiliation{Centre for Cosmology, Particle Physics and Phenomenology - CP3, Universit{\'e} catholique de Louvain, Louvain-la-Neuve, Belgium}
\affiliation{Department of Physics, Mercer University, Macon, GA 31207-0001, USA}
\affiliation{Dept. of Astronomy, University of Wisconsin{\textendash}Madison, Madison, WI 53706, USA}
\affiliation{Dept. of Physics and Wisconsin IceCube Particle Astrophysics Center, University of Wisconsin{\textendash}Madison, Madison, WI 53706, USA}
\affiliation{Institute of Physics, University of Mainz, Staudinger Weg 7, D-55099 Mainz, Germany}
\affiliation{Department of Physics, Marquette University, Milwaukee, WI, 53201, USA}
\affiliation{Institut f{\"u}r Kernphysik, Westf{\"a}lische Wilhelms-Universit{\"a}t M{\"u}nster, D-48149 M{\"u}nster, Germany}
\affiliation{Bartol Research Institute and Dept. of Physics and Astronomy, University of Delaware, Newark, DE 19716, USA}
\affiliation{Dept. of Physics, Yale University, New Haven, CT 06520, USA}
\affiliation{Dept. of Physics, University of Oxford, Parks Road, Oxford OX1 3PU, UK}
\affiliation{Dept. of Physics, Drexel University, 3141 Chestnut Street, Philadelphia, PA 19104, USA}
\affiliation{Physics Department, South Dakota School of Mines and Technology, Rapid City, SD 57701, USA}
\affiliation{Dept. of Physics, University of Wisconsin, River Falls, WI 54022, USA}
\affiliation{Dept. of Physics and Astronomy, University of Rochester, Rochester, NY 14627, USA}
\affiliation{Department of Physics and Astronomy, University of Utah, Salt Lake City, UT 84112, USA}
\affiliation{Oskar Klein Centre and Dept. of Physics, Stockholm University, SE-10691 Stockholm, Sweden}
\affiliation{Dept. of Physics and Astronomy, Stony Brook University, Stony Brook, NY 11794-3800, USA}
\affiliation{Dept. of Physics, Sungkyunkwan University, Suwon 16419, Korea}
\affiliation{Institute of Basic Science, Sungkyunkwan University, Suwon 16419, Korea}
\affiliation{Institute of Physics, Academia Sinica, Taipei, 11529, Taiwan}
\affiliation{Dept. of Physics and Astronomy, University of Alabama, Tuscaloosa, AL 35487, USA}
\affiliation{Dept. of Astronomy and Astrophysics, Pennsylvania State University, University Park, PA 16802, USA}
\affiliation{Dept. of Physics, Pennsylvania State University, University Park, PA 16802, USA}
\affiliation{Dept. of Physics and Astronomy, Uppsala University, Box 516, S-75120 Uppsala, Sweden}
\affiliation{Dept. of Physics, University of Wuppertal, D-42119 Wuppertal, Germany}
\affiliation{DESY, D-15738 Zeuthen, Germany}

\author[0000-0001-6141-4205]{R. Abbasi}
\affiliation{Department of Physics, Loyola University Chicago, Chicago, IL 60660, USA}

\author[0000-0001-8952-588X]{M. Ackermann}
\affiliation{DESY, D-15738 Zeuthen, Germany}

\author{J. Adams}
\affiliation{Dept. of Physics and Astronomy, University of Canterbury, Private Bag 4800, Christchurch, New Zealand}

\author[0000-0003-2252-9514]{J. A. Aguilar}
\affiliation{Universit{\'e} Libre de Bruxelles, Science Faculty CP230, B-1050 Brussels, Belgium}

\author[0000-0003-0709-5631]{M. Ahlers}
\affiliation{Niels Bohr Institute, University of Copenhagen, DK-2100 Copenhagen, Denmark}

\author{M. Ahrens}
\affiliation{Oskar Klein Centre and Dept. of Physics, Stockholm University, SE-10691 Stockholm, Sweden}

\author[0000-0002-9534-9189]{J.M. Alameddine}
\affiliation{Dept. of Physics, TU Dortmund University, D-44221 Dortmund, Germany}

\author{A. A. Alves Jr.}
\affiliation{Karlsruhe Institute of Technology, Institute for Astroparticle Physics, D-76021 Karlsruhe, Germany }

\author{N. M. Amin}
\affiliation{Bartol Research Institute and Dept. of Physics and Astronomy, University of Delaware, Newark, DE 19716, USA}

\author{K. Andeen}
\affiliation{Department of Physics, Marquette University, Milwaukee, WI, 53201, USA}

% ...

\date{\today}

\collaboration{382}{IceCube Collaboration}

\begin{abstract}
Galaxy clusters have the potential to accelerate cosmic rays (CRs) to ultra-high energies via accretion shocks or embedded CR acceleration sites.
CRs with energies below the Hillas condition will be confined within the cluster and will eventually interact with the intracluster medium (ICM) gas to produce secondary neutrinos and $\gamma$ rays.
Using 9.5 years of muon-neutrino track events from the IceCube Neutrino Observatory, we report the results of a stacking analysis of 1094 galaxy clusters, with masses $\gtrsim 10^{14}$ \(\textup{M}_\odot\) and redshifts between 0.01 and $\sim$1, detected by the {\it Planck} mission via the Sunyaev-Zeldovich (SZ) effect.
We find no evidence for significant neutrino emission and report upper limits on the cumulative unresolved neutrino flux from massive galaxy clusters after accounting for the completeness of the catalog up to a redshift of 2, assuming three different weighting scenarios for the stacking and three different power-law spectra. Weighting the sources according to mass and distance, we set upper limits at $90\%$ confidence level that constrain the flux of neutrinos from massive galaxy clusters ($\gtrsim 10^{14}$ \(\textup{M}_\odot\)) to be no more than $4.6\%$ of the diffuse IceCube observations at 100~TeV, assuming an unbroken $E^{-2.5}$ power-law spectrum.

\end{abstract}



\section{Introduction} \label{sec:intro}

The IceCube Neutrino Observatory has been observing a steady diffuse flux of TeV--PeV astrophysical neutrinos for over a decade \citep{diffuse}. Despite recent indications of potential point sources of neutrinos ~\citep{txs,txs2,ngc}, the origin of the majority of the flux remains undiscovered. The mystery is closely tied to the sources of ultra-high-energy ($> 10^{18}$ eV) cosmic rays (UHECRs) ~\citep{2013ApJ...768L...1A,PierreAuger:2016use}, as well as the GeV--TeV extra-galactic gamma-ray background (EGB) observed by {\it Fermi}-LAT \citep{Fermi-LAT:2015qzw} -- both of which have been observed to produce energy fluxes comparable to that of high-energy neutrinos above 100 TeV. This has led several authors to propose a unique astrophysical source class that could explain all three aforementioned fluxes (e.g., \cite{fangmurase2018,2015A&A...578A..32Z,Murase:2008yt,Tamborra:2014xia}). Galaxy clusters are one such viable class of sources.       

During large-scale structure formation, clusters of galaxies emerge as a result of accretion and mergers of smaller structures, producing Mpc-scale shock waves~\citep{miniati2000,ryu2003} and a turbulent intra-cluster medium (ICM). The shocks can potentially accelerate CRs to energies as high as $10^{20}$~eV~\citep{inoue}, depending on the mass of the cluster~\citep{fangolinto2016, 2011A&A...527A..99E,Wiener:2018sya}, through the first order Fermi acceleration mechanism~\citep{blandford}. Other proposed candidates for CR acceleration within clusters are  
embedded compact objects such as active Active Galactic Nuclei (AGN) ~\citep{stecker1991,winter2013,fangmurase2018} and magnetars ~\citep{murase09}. 

The predicted maximum accelerated CR energy and resultant neutrino flux are highly dependent on the properties of the embedded sources, and vary greatly from model to model.
For both the Mpc-scale shock and embedded source scenarios, the accelerated CRs are confined in clusters for times exceeding the age of the universe~\citep{berezinsky1997}.
These CRs interact with the ICM to produce pions, which decay, yielding neutrinos and $\gamma$ rays. 

Galaxy clusters remain unobserved in very-high-energy gamma rays ($>100$~GeV). {\it Fermi}-LAT \citep{2013MNRAS.429.2069D,2013A&A...560A..64H} as well as Imaging Air Cherenkov Telescopes (IACTs;  \cite{2012ApJ...757..123A,MAGIC:2018tuz,2012ApJ...750..123A}) have reported upper limits on the gamma-ray emission from several nearby clusters. A detection of astrophysical neutrinos from galaxy clusters would constitute incontrovertible evidence for proton acceleration in these objects. Previous IceCube analyses have reported constraints on the neutrino emission from five nearby clusters~\citep{IC40_cluster,IC40_cluster2}. However, the source class as a whole remains a largely uncharted territory in terms of neutrino searches. Some works~\citep{2021MNRAS.507.1762H, fangmurase2018} predict that almost the entirety of the diffuse flux observed by IceCube could be explained by the cumulative emission from galaxy clusters. In this work, for the first time, we use cosmological data to test the contribution of galaxy clusters to the diffuse astrophysical neutrino flux. In particular, we focus on testing the scenarios in which the most massive clusters accelerate CRs in accretion shocks and/or cluster mergers \citep{fangolinto2016,2021MNRAS.507.1762H,Hussain:2022tls}.

\section{Galaxy Cluster Catalog}
\label{sec:catalog}
The mass and redshift range of galaxy clusters with the largest contribution to the neutrino flux varies across models. For this study, we choose a catalog of clusters that extends in redshift coverage between 0.01 and 0.97, and has cluster masses above $7.8 \times 10^{13} \textup{ M}_{\odot}$, where $\textup{M}_{\odot}$ is the solar mass. This selection allows us to test neutrino production in some of the largest and latest-formed objects in the universe.           
Our catalog of galaxy clusters is taken directly from the {\it Planck} 2015 survey~\citep{planck_2015}.
The catalog consists of all galaxy clusters detected through the variations observed in the brightness of cosmic microwave background (CMB). These variations are caused by the inverse Compton scattering of CMB photons off high-energy electrons in the ICM, known as the Sunyaev-Zeldovich (SZ) effect~\citep{2002ARA&A..40..643C}. In this work, we solely use the 1094 sources for which a redshift estimate is available. The resulting selection of clusters covers 83\% of the sky. 


\subsection{Catalog Completeness}
\label{sec:catalog:comp}
In order to extrapolate the results from our source sample to the entire population of galaxy clusters, we need an estimate for the completeness of the {\it Planck} catalog. This requires a calculation of the fraction of clusters that would be reliably detected by the {\it Planck} mission given a theoretical distribution of galaxy clusters in the universe. 
We extract {\it Planck}'s completeness as a function of redshift and cluster mass from the information given in~\citet{planck_2015} using simulations. We describe the steps briefly here and point the reader to \cite{raghuanthan21} for more details.

We start by simulating maps in individual {\it Planck} frequency bands which contain signals from astrophysical foregrounds, the CMB, instrumental noise, and the thermal Sunyaev-Zeldovich (tSZ) signal. The astrophysical foregrounds include contributions from diffuse SZ signals (both kinematic and thermal) and also emission from dusty star-forming and radio galaxies. Galactic foregrounds are ignored since the {\it Planck} galaxy-cluster catalog is constructed after applying a mask to remove the majority of the galactic dust emission. The tSZ signal from a hypothetical cluster of given mass and redshift is modeled using a generalized Navarro-Frenk-White (NFW) profile \citep{nagai07} calibrated with X-ray observations by \citet{arnaud09}. The dimensionless pressure profile of the ICM is integrated along the line of sight to obtain the Compton-y signal $y(x)$, where $x$ is a given position on the sky. 

Next, we integrate $y(x)$ over the angular extent of the cluster, $R_{500c}$, to obtain the integrated Comptonization parameter, $Y_{500c}$. Here the subscript $500c$ refers to a sphere where the mean mass over-density of the cluster is at least 500 times the cosmic critical density at the redshift of the cluster. $Y_{500c}$ is directly related to the cluster mass \citep{plancksz15}. 

\begin{figure}[h]
\includegraphics[width=0.99\textwidth]{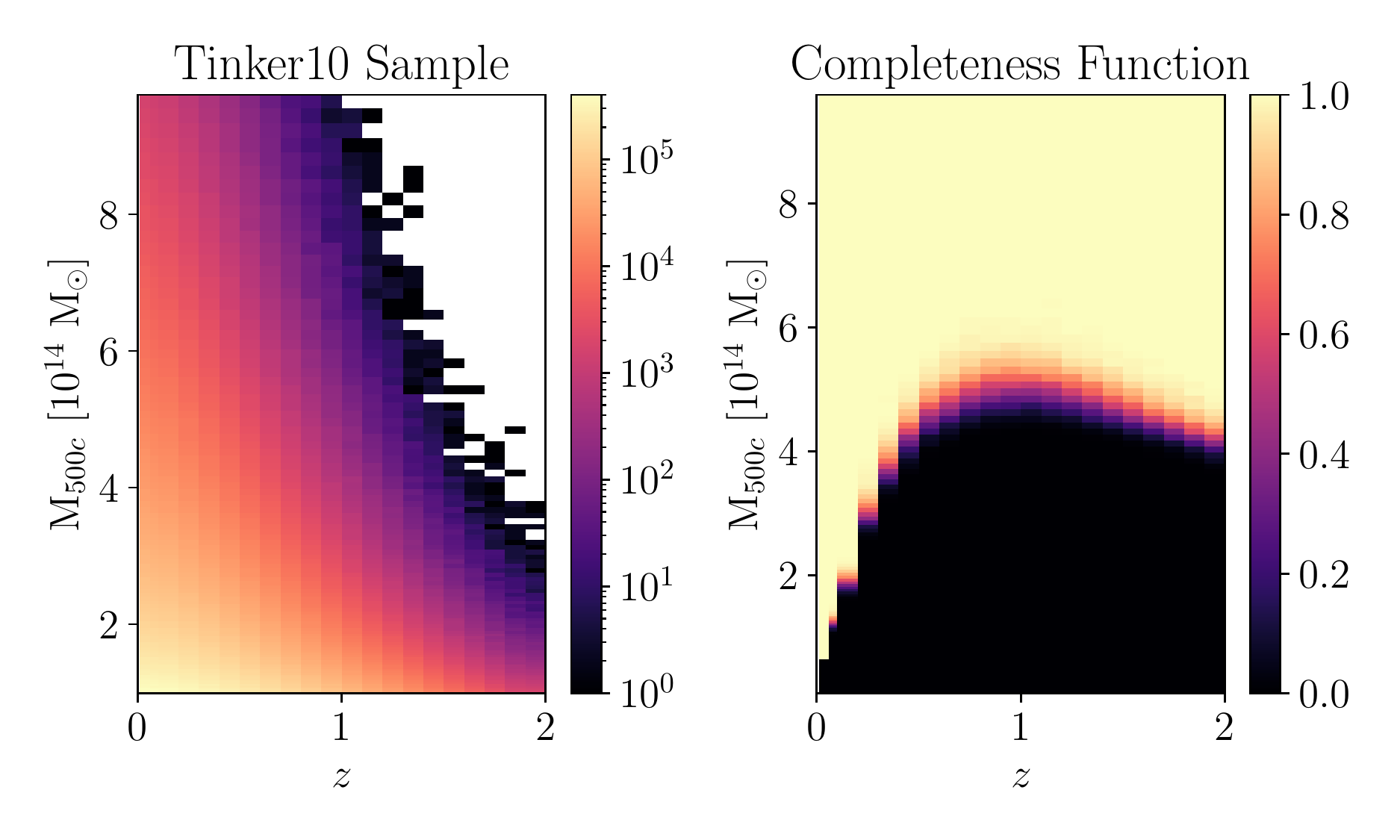}
\caption{Left: The theoretical sample of $\mathcal{O}(10^5)$ clusters drawn from the Tinker 2010 hmf, binned in redshift and mass. Right: The completeness function for the {\it Planck} catalog considered in this work. The theoretical sample is scaled with the completeness function to calculate the completeness of the catalog (see text).}
\label{fig:fig0}

\end{figure}

Following the maximum likelihood approach described in \cite{raghuanthan21}, we combine the simulated maps from all {\it Planck} frequency bands and obtain the signal-to-noise ratio (SNR) for each cluster as a function of mass ($\mvir$) and redshift ($z$) in the ranges: ${\rm log}(\mvir/\msol) \in [13,16]$ in logarithmic steps of 0.1 and $z \in [0.1, 3]$ in steps of 0.1. For detection, a cluster must have a minimum SNR of 5. Using the detected clusters, we estimate the catalog completeness as \citep{plancksz15, alonso16},
\begin{equation}
\chi(\Yvir) = \frac{1}{2} \left[ 1 + {\rm erf} \left( \frac{\Yvir^{\rm true} - q_{\rm lim} \sigma_{\Yvir}} {\sqrt{2}\sigma_{\Yvir}}\right)\right],
\end{equation}
where $q_{\rm lim} = \snrlimit$ is the detection threshold, $\sigma_{\Yvir}$ is the measurement uncertainty of the integrated $\Yvir$ signal estimated above in simulations, and $\Yvir^{\rm true}$ is the true $\Yvir$ flux. Figure \ref{fig:fig0} (right panel) shows the completeness obtained as a function of redshift and mass.

We then create a theoretical distribution of galaxy clusters by drawing $\mathcal{O}(10^5)$ samples from the Tinker 2010 halo mass function (hmf) ~\citep{tinker_2010} binned in redshift and mass (Figure \ref{fig:fig0}, left panel).  We only consider galaxy clusters with masses between $10^{14}$ \(\textup{M}_\odot\) and $10^{15}$ \(\textup{M}_\odot\) and  redshifts between 0.01 and 2 in our simulation. In our energy range of interest, and in models where CR acceleration takes place through shock waves due to the ICM accretion onto clusters, clusters with masses below $10^{14}$ \(\textup{M}_\odot\) or $z>1$ are not expected to produce a significant flux of neutrinos at earth \citep{fangolinto2016}. In the aforementioned scenarios, the bulk of the flux contribution comes from $0.3 < z < 1$ in accordance with the cosmological distribution of sources \citep{fangolinto2016}. 

We scale this theoretical distribution with the {\it Planck} completeness function obtained above to estimate the fraction of detectable clusters. This fraction is determined by dividing the sum of the scaled distribution by the total number of clusters in the hmf sample. In the case where each redshift and mass bin is equally weighted,
we find a catalog completeness of $\sim 17 \%$. Other weighting schemes are discussed in the following section. We note that our completeness calculations are robust within $2 \%$ under different choices of the hmf.

\subsection{Source Weighting}
\label{sec:catalog:w}
In this work, we consider three scenarios in which the properties of galaxy clusters may be correlated with their neutrino flux on earth. These scenarios are encoded in the weighting schemes used in the stacking analysis (Section \ref{sec:analysis}). We assign each cluster a weight $w_k$ where the index $k$ runs over all clusters in the catalog.

The simplest and most conservative scenario is that of equal weights, where we set   
$w_{k} = 1$ for all clusters. Here we effectively assign equal relative neutrino intensity to each object in the catalog, regardless of its properties.
We note that this is an unphysical scenario, and results based on this weighting scheme represent the most conservative upper bounds obtained with this analysis.

A more realistic scenario is one where the observed neutrino flux from a cluster is inversely proportional to the cluster's distance squared. In this weighting scheme, we assign weights as $w_{k} = d_{k}^{-2}$, where $d_{k}$ is the cluster distance obtained from its redshift using the cosmological parameters from {\it Planck}~\citep{planck_2015}. We refer to this as ``distance weighting'' throughout the paper.

In addition to a flux that scales with $d_{k}^{-2}$, more massive clusters with longer formation histories are expected to have a higher CR content \citep{berezinsky1997}, with the    
CR luminosity being proportional to the cluster mass 
~\citep{fangolinto2016}. 
Our final scenario weights sources by $w_{k} = M_{k}d_{k}^{-2}$, where $M_{k}$ is the source mass.
We will refer to this weighting scheme as ``mass weighting'' throughout this paper. 

Finally, for each weighting scheme, we calculate an ``effective completeness'' of the catalog by weighting the mass and redshift bins following the procedure described in Section~\ref{sec:catalog:comp}. The distance weighting and the mass weighting have an estimated effective completeness of $\sim53 \%$, and $\sim66 \%$, respectively.

We note two caveats with respect to our effective completeness calculation. Our calculation does not extend to scenarios in which the bulk of CR acceleration is driven by sources embedded in less massive ($<10^{14} $ \(\textup{M}_\odot\) clusters at high $z$. For example, reducing the minimum halo mass in our simulations from $10^{14}$ \(\textup{M}_\odot\) to $5 \times 10^{13}$ \(\textup{M}_\odot\) would decrease the mass-weighted completeness to $36\%$. In addition, considering a cosmological evolution of source density proportional to $(1+z)^3$ \citep{fangmurase2018} by introducing an additional weighting factor in each redshift bin, would reduce the distance weighted completeness to $0.5\%$ and mass-weighted completeness to $2\%$. (X-ray AGN typically evolve with $(1+z)^{4.8}$, in which the completeness is further reduced.) Models of diffuse neutrino production through large-scale structure formation are less sensitive to the contribution from low-mass halos or $z > 1$ and are therefore suitably tested in our analysis.

\section{Detector and Dataset}
\label{sec:detector}

The IceCube Neutrino Observatory instruments a cubic kilometer of South Pole ice with 5160 digital optical modules (DOMs) deployed between 1450~m and 2450~m below the surface in Antarctica.
IceCube is composed of 86 vertical strings, with 60 DOMs per string. 
The DOMs consist of a photomultiplier tube (PMT) and on-board readout electronics designed to detect Cherenkov light induced by charged particles that can be produced in charged-current (CC) and neutral-current (NC) neutrino interactions~\citep{icecube_daq,icecube_detector}. 
These secondary particles result in either track-like (CC interactions of muon neutrinos) or cascade-like (all other CC and NC interactions) optical signatures that are recorded by DOMs. The photon counts and timing information are used to reconstruct both the energy deposition and particle direction.
IceCube can detect all flavors of neutrinos and can't discriminate between neutrinos and anti-neutrinos. The angular resolution achieved depends on the optical signature, with tracks yielding the best resolution, typically below 1$^\circ$ above 1 TeV~\citep{icecube_7year}.

Due to their superior angular resolution, the analysis presented in this work uses a 9.5-year data sample consisting exclusively of muon tracks.
The data set, recorded between April 2008 and November 2017~\citep{icecube_7year,realtime}, covers the full sky.  
The majority of the dataset consists of background events induced by atmospheric neutrinos and atmospheric muons created by CR air showers in the upper atmosphere. A detailed description of the dataset and the related reconstruction algorithms used in this work can be found in~\citet{IceCube:2013dkx,IceCube:2010whx}.

\section{Analysis}
\label{sec:analysis}
\subsection{Likelihood and Test Statistic}

In order to search for the cumulative neutrino emission from our catalog of galaxy clusters in data dominated by atmospheric backgrounds, we perform a stacking analysis using an unbinned maximum-likelihood method~\citep{unbinned_llh, IceCube:2006flj}. 

The likelihood function for $M$ stacked clusters is defined as,

\begin{equation}
\mathcal{L} \left( n_{s}, \gamma \right) = \prod_{i}^{N} \left( \frac{n_{s}}{N} \sum_{k}^{M} \frac{r_k w_k}{\sum_{j}^{M} r_j w_j} S_{i}^{k} \left( \gamma \right) + \left( 1 - \frac{n_{s}}{N} \right) B_{i} \right),
\label{eq:llh}
\end{equation}
where $n_{s}$ is the number of signal events, $\gamma$ is the spectral index of an assumed unbroken power-law spectrum that is common to all sources, $N$ is the total number of data events, $S_{i}^{k}$ is the signal probability distribution function (PDF) for the $i$th neutrino event for the $k$-th source, $B_{i}$ is the background PDF for the $i$-th neutrino event, and $r_{k}$ and $w_k$ are stacking weights described below. 

The background PDF is a function of both the reconstructed event energy, $E_{i}$, and the reconstructed event declination, $\delta_{i}$, and is calculated by randomizing the right ascension of events in a given declination band. The signal PDF is a weighted superposition of the signal PDFs of all sources in our selection of galaxy clusters. A potential source of astrophysical neutrinos should be distinguishable from background due to the spatial clustering of high-energy events in its vicinity. Therefore, each $S_{i}^{k}$ consists of a Gaussian spatial term and an energy term that favors high-energy events,
\begin{equation}
S_{i}^{k} = \frac{1}{2\pi \sigma_i^2} \exp \left ( - \frac{|\Psi_{ik}|^2}{2\sigma_i^2} \right ) \times \mathcal{E} (E_i,\delta_i,\gamma).
\end{equation}
The term $\Psi_{ik}$ is the angular separation between event $i$ and source $k$, $\sigma_{i}$ \footnote{Only the event angular uncertainty, $\sigma_{i}$, is considered because the source angular uncertainty, $\sigma_{k}$, is known to much higher accuracy. {\it Planck's} angular resolution is on the order of 1~arcmin~\citep{planck_2015}, while IceCube's angular resolution is energy dependent, but at least an order of magnitude less accurate~\citep{icecube_7year}.} is the angular uncertainty of event $i$, and $\mathcal{E} (E_i,\delta_i,\gamma)$ is a simulated energy PDF, where $\gamma$ is assumed to have a value between 1 and 4. Nearby clusters with extended profiles are not part of our sample, therefore the likelihood function is independent of source extension. 

The source weight $w_k$ in Eq.~\ref{eq:llh} parameterizes the relative contribution of each cluster to the total signal. It is determined by the weighting schemes described in section~\ref{sec:catalog:w}. 
The model-independent term, $r_k$, depends on the detector acceptance, and is determined by the effective area of the detector at the source location and the assumed neutrino spectrum. 
For more details see \citet{IceCube:2021waz,IceCube:2016qvd}.

In order to determine the number of signal events and the spectral index that best describe the data, the likelihood in Eq.~\ref{eq:llh} is maximized with respect to $n_s$ and $\gamma$, and tested against the null hypothesis, which  corresponds to the background-only scenario ($n_s = 0$). For that purpose, we define a test statistic (TS) as,  
\begin{equation}
\mathrm{TS} = 2\log\frac{\mathcal{L}\left( \hat{n}_{s}, \hat{\gamma}\right)}{\mathcal{L} \left( n_{s} = 0 \right)},    
\end{equation}
where $\hat{n}_{s}$ and $\hat{\gamma}$ are the best-fit values. We note that each choice of weights will result in a different fitted $\hat{n}_s$, $\hat{\gamma}$ and TS.
\subsection{Sensitivity}
\label{sec:sensi}
Before looking at the actual data, we performed an ensemble of pseudo-experiments to determine the sensitivity of the analysis. 
First, we generated a set of background-only pseudo-experiments by randomizing the data in right ascension. For each such scramble, we used the maximum-likelihood method and calculated a corresponding TS. 
This allowed us to create a background-only TS distribution with 20,000 scrambles.  

We then constructed hundreds of 
simulations where we injected pseudosignal events into the scrambled data. For a fixed number of injected events, we performed several hundreds of pseudo-experiments to determine the corresponding TS distribution. This process was repeated for increased numbers of pseudosignal events. We simulated the pseudosignal neutrinos according to a power-law spectrum with a given spectral index $\gamma \in \lbrace 2.0, 2.5, 3.0 \rbrace$, and injected them at the locations of the galaxy cluster sources. The number of events injected at a given source position is weighted by the corresponding stacking weight of that source.
The number of injected signal events for which the TS values exceed the median of the background-only TS distribution 90\% of the time is defined as the sensitivity of the analysis for an astrophysical signal originating from galaxy clusters. Alternatively, this sensitivity corresponds to the median upper limit at 90\% confidence level (CL) that is set in an ensemble of background-only pseudo-experiments. We find that for a given weighting scheme, this analysis is most sensitive to a hard spectral index of 2.0. This is expected as a hard spectrum is more distinguishable from the atmospheric background which follows a significantly softer spectrum \citep{IceCube:2010whx}.

\section{Results}
\label{sec:results}

We perform a stacking analysis on our catalog as described in Section~\ref{sec:analysis} for the different weighting schemes described in Section~\ref{sec:catalog:w}.
None of the scenarios we investigate show a significant excess of signal events over background. The best-fit number of signal events, as well as the corresponding TS value, is zero for each weighting scheme.
 
In order not to report overly constraining limits due to potential under-fluctuations in the data, we set the upper limits at the 90\% sensitivity level described above. Table \ref{tab:contrib} shows the upper limits on the unresolved flux of galaxy clusters at 100 TeV for each weighting scenario after scaling for the effective completeness of the catalog and correcting for the 83\% sky coverage of {\it Planck}. 

\begin{figure}[t]
\includegraphics[width=\textwidth]{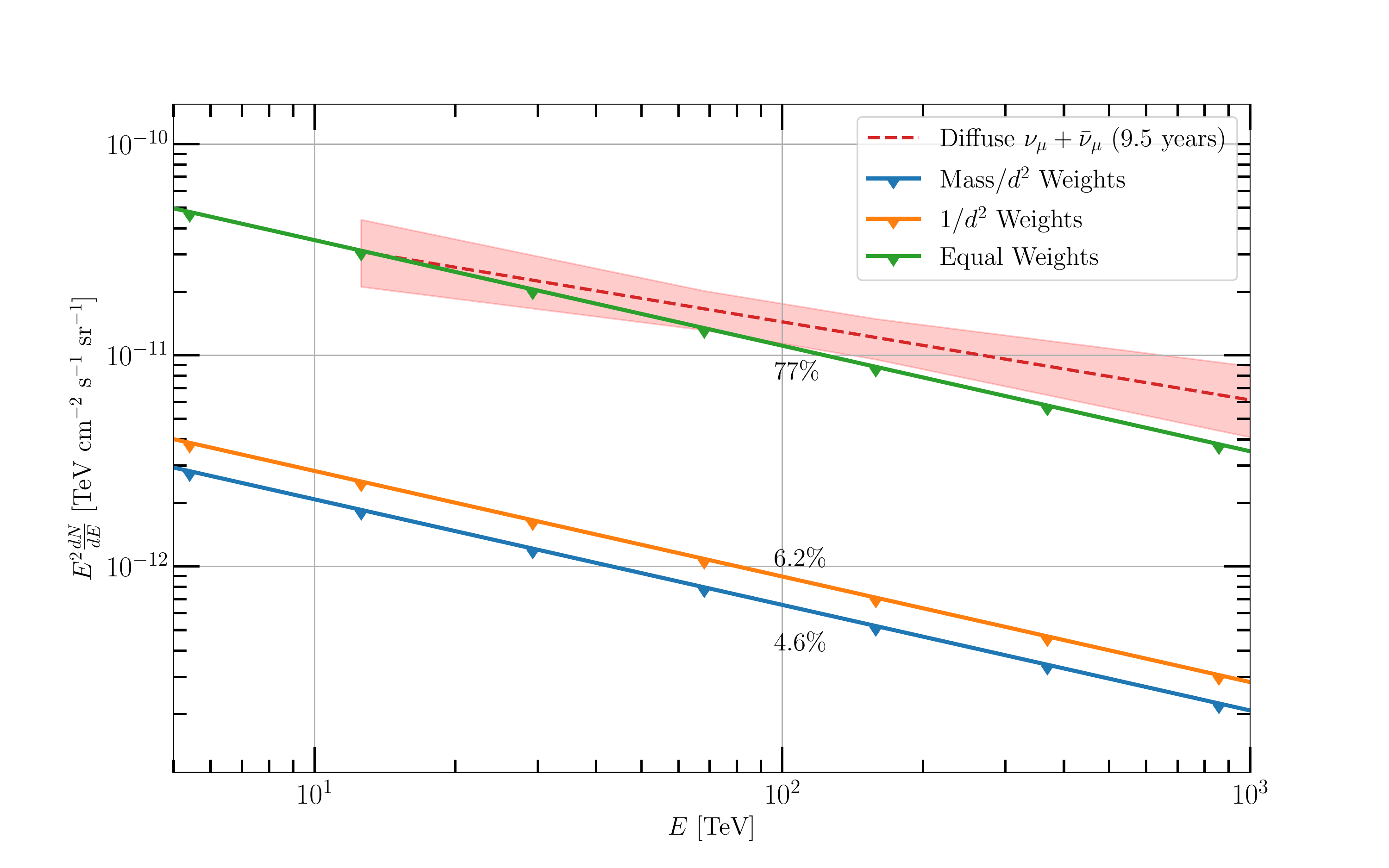}
\caption{IceCube limits at 90\% CL on the diffuse muon-neutrino flux from galaxy clusters in the {\it Planck} SZ Catalog~\citep{planck_2015} for an unbroken $E^{-2.5}$ power-law spectrum. The limits have been scaled for the completeness of the catalog for each weighting scheme considered (see text) and are plotted for the central 90\% of signal energies that contribute to the sensitivity. We exclude galaxy clusters from the {\it Planck} SZ catalog as the sole sources responsible for the diffuse muon neutrino observations in \citet{IceCube:2021uhz}, for all weighting schemes considered, with the fractional contribution of galaxy clusters at 100 TeV indicated in the figure (77\% , 6.2\% and 4.6\% for equal, distance and mass weights respectively). Constraints for other spectra are presented in Table~\ref{tab:contrib}.}
\label{fig:fig2}
\end{figure}

The upper limits set in this analysis can then be compared with the diffuse muon-neutrino flux measured by IceCube~\citep{IceCube:2021uhz} to estimate the maximum contribution of the galaxy cluster population to the diffuse observations. This fraction is also reported in Table~\ref{tab:contrib} for each scenario and assumed spectral index. In addition, Figure~\ref{fig:fig2} shows the upper limits on the neutrino emission from clusters for an assumed $E^{-2.5}$ power-law spectrum. This limit is directly compared to the diffuse muon-neutrino observations over the full sensitive energy range of IceCube, which have a best fit $E^{-2.37}$ power-law spectrum~\citep{IceCube:2021uhz}.

%
  
We find the equal-weighting case to provide the weakest constraints across all spectral indices.
This is expected, because this weighting scheme is meant to act as a check on our methods, without corresponding to a realistic, physical scenario.
The distance weighting and the mass weighting schemes provide similar constraints to one another, which is also expected, as they are similar in terms of both the physics cases they are testing and their effective completeness. Both cases assume that the neutrino flux falls off with the cluster distance as $d^{-2}$. Distance weighting assumes that galaxy clusters are neutrino standard candles. On the other hand, the mass weighting assumes linear proportionality between neutrino luminosity and cluster mass. For a spectral index of 2.5

the results exclude a maximum contribution of 77\% at 100 TeV from galaxy clusters assuming equal weights. Under the more realistic distance and mass weighting schemes, this contribution is constrained down to 6.2\% and 4.6\% at 100 TeV, respectively.

   \begin{deluxetable*}{c| c c |c c| c c}
     \tabletypesize{\small}
    \tablecaption{Upper limits at 90\% CL on the diffuse muon-neutrino flux of {\it Planck} SZ galaxy clusters, after scaling for the effective catalog completeness corresponding to each weighting scheme considered. For each weighting scheme, upper limits are quoted for three $E^{-\gamma}$ power-law spectra. The upper limits are presented in terms of the diffuse galaxy-cluster flux at E$_0$ = 100 TeV, $E_0^2 \frac{dN}{dE}$, as well as its relative contribution to the diffuse muon-neutrino observations of \citet{IceCube:2021uhz} at 100 TeV. All fluxes are given in units of TeV cm$^{-2}$ s$^{-1}$ sr$^{-1}$.
      \label{tab:contrib}}
    \tablehead{
      \colhead{} & \multicolumn{2}{c}{$\gamma = 2.0$} & \multicolumn{2}{c}{$\gamma = 2.5$} & \multicolumn{2}{c}{$\gamma = 3.0$}\\
     \cline{2-7}
      \colhead{Weighting Scheme} & \colhead{$E_0^2 \frac{dN}{dE}$} & \colhead{Diffuse Contribution} &  \colhead{$E_0^2 \frac{dN}{dE}$} & \colhead{Diffuse Contribution}  & \colhead{$E_0^2 \frac{dN}{dE}$} & \colhead{Diffuse Contribution}
       }
    \startdata
      Equal & $5.31 \times 10^{-12}$ & 37\% & $1.11 \times 10^{-11}$ & 77\% & $4.70 \times 10^{-12}$ & 32.7\%  \\
 Distance   & $6.53 \times 10^{-13}$ &  4.5\%  & $8.96 \times 10^{-13}$ & 6.2\% &$2.41 \times 10^{-13}$ & 1.7\%\\
 Mass  & $6.12 \times 10^{-13}$ & 4.3\% & $6.58 \times 10^{-13}$ & 4.6\% &  $2.21 \times 10^{-13}$ & 1.5\% 
    \enddata
  \end{deluxetable*}%


\begin{figure}[ht]
\includegraphics[width=\textwidth]{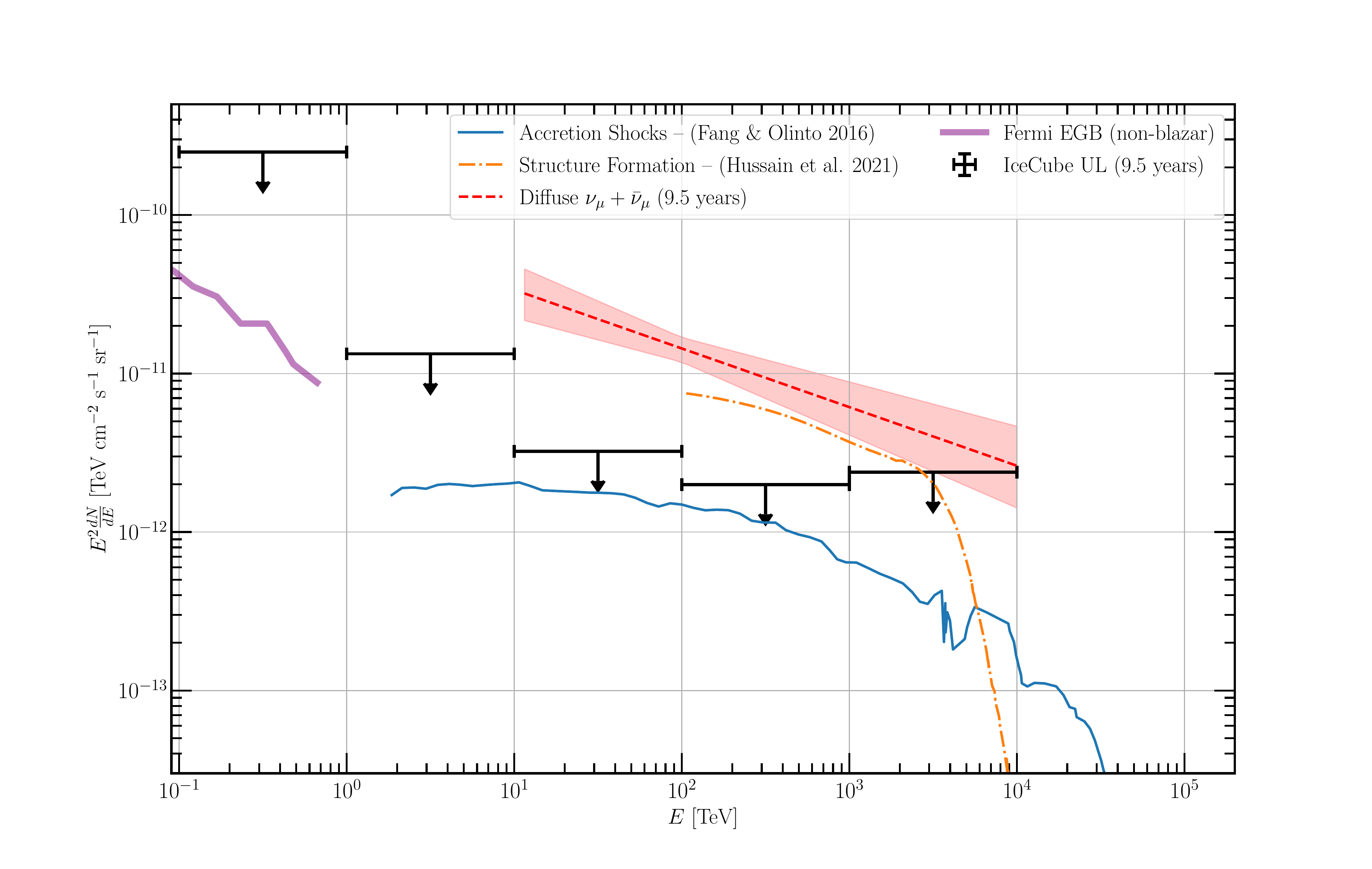}
\caption{Quasi-differential upper limits at 90\% CL on the total neutrino flux from galaxy
clusters in one-decade energy bins for the distance
weighting scheme. The limits have been scaled to account for the completeness
of the catalog. The blue
solid line shows the model prediction from \citet{fangolinto2016} in which accretion shocks in
galaxy clusters are responsible for CR acceleration. The orange dashed-dotted line shows
the total neutrino prediction from an updated MHD simulation of the aforementioned scenario \citep{2021MNRAS.507.1762H} assuming a CR injection spectrum of 1.5, a kinetic energy conversion rate of $0.5\%$ and no source evolution. The non-blazar contribution to the extragalactic gamma-ray background (EGB) observed by {\it Fermi}-LAT \citep{Fermi-LAT:2014ryh} is shown as the thick purple line. The IceCube diffuse muon-neutrino observations of \citet{IceCube:2021uhz} are indicated by the red dashed line and the red band.}
\label{fig:diffsensi}
\end{figure}

 
Another useful way of comparing the contribution of galaxy clusters to diffuse neutrino observations is by evaluating the upper limits in quasi-differential energy bins. We calculate the  upper limits in one-decade energy bins, assuming an $E^{-2}$ spectrum across the bin. 
Figure~\ref{fig:diffsensi} shows the quasi-differential limits for the distance weighting scheme along with comparable theoretical predictions from the mass and redshift range used in this work. 
The IceCube upper limits obtained in this work are most constraining between 10 TeV and 1 PeV compared to the diffuse neutrino observations, where we find that the population of galaxy clusters in our mass and redshift range of interest cannot contribute to more than 9\%--13\% of the diffuse neutrino flux. 

Our results above 100 TeV constrain the model based on magnetohydrodynamic (MHD) simulations by \citet{2021MNRAS.507.1762H}, which predicts that most of the contribution to the diffuse neutrino flux can be attributed to clusters with masses above $10^{14}~\textup{M}_{\odot}$. For the aforementioned model, our limits exclude a CR injection spectrum of 1.5, and a conservative energy conversion rate of 0.5\%. Our limits are robust  between 100 TeV and 1 PeV to the choice of the injection spectrum and maximum CR-energy. These conclusions can change for a different choice of kinetic energy conversion rates and assumptions about source evolution.  
We note that certain models of structure formation followed by hadronic interactions in galaxy clusters suggest that the total flux from clusters is indeed very low and not able to explain more than a few percent of the total diffuse flux \citep{2015A&A...578A..32Z,fangolinto2016}. Our analysis is not sensitive enough to probe such models. As explained in section \ref{sec:catalog:w}, models that assume AGN as the embedded sources of CRs in clusters \citep{fangmurase2018} are not constrained using the mass range of clusters tested in this work.

\section{Conclusions}
\label{sec:conclusions}
In this work, we presented a search for neutrinos produced by a large catalog of galaxy clusters, which are expected to produce neutrinos through embedded sources, accretion shocks, or both. 
We performed a stacking analysis with 1094 SZ-selected galaxy clusters from {\it Planck} using 9.5 years of muon-track IceCube data. Three different weighting schemes were investigated, i.e., equal weights, distance weights ($1/d^2$), and mass weights ($M/d^2$).
We find no evidence for neutrino emission in any of the scenarios considered. Assuming an unbroken $E^{-\gamma}$ power-law flux common among all galaxy clusters, we compute upper limits at 90\% CL on the diffuse neutrino flux from galaxy clusters with masses between $10^{14}$ \(\textup{M}_\odot\) and $10^{15}$ \(\textup{M}_\odot\) and a redshift between 0.01 and 2 using a (simulated and extrapolated in $z$) complete catalog of galaxy clusters. We constrain the galaxy-cluster contribution to the diffuse muon-neutrino observations of \cite{IceCube:2021uhz} at 100 TeV.  In particular, for $\gamma = 2.5$, galaxy clusters can contribute no more than 67\%, 6.2\%, and 4.6\% for the equal, distance, and mass weighting scenarios, respectively.

We also compute quasi-differential limits at 90\% CL on the diffuse neutrino flux from galaxy clusters. These limits exclude the MHD model of \citet{2021MNRAS.507.1762H} as neutrino-production mechanisms in galaxy clusters. Accretion-shock model predictions by \citet{2015A&A...578A..32Z,fangolinto2016} are not constrained by our limits, making our results consistent with scenarios in which accretion shocks make a sub-dominant contribution to the diffuse neutrino flux \citep{Murase:2016gly}. 
   
An absence of GeV--TeV gamma rays from nearby clusters, along with our new IceCube limits, constrains the shock-acceleration efficiencies in galaxy clusters. Galaxy clusters have long been proposed as the single unified solution to explain the diffuse fluxes of UHECRs, neutrinos, and gamma rays~\citep{fangmurase2018}. Our results encourage a revisiting of the multi-messenger connections between different classes of these high-energy particles. The next generation of observatories such as IceCube Gen-2 \citep{IceCube-Gen2:2020qha}, CTA \citep{Perez-Romero:2021gxh} and AMEGO-X \citep{Fleischhack:2021mhc}, offers a promising prospect for the detection of smoking-gun MeV-PeV multimessenger signatures of CR acceleration in galaxy clusters.

\acknowledgments

The IceCube Collaboration acknowledges the significant contributions made to this manuscript by Mehr Un Nisa, Andrew Ludwig and Srinivasan Raghunatha. We also acknowledge support from: 
USA {\textendash} U.S. National Science Foundation-Office of Polar Programs,
U.S. National Science Foundation-Physics Division,
U.S. National Science Foundation-EPSCoR,
Wisconsin Alumni Research Foundation,
Center for High Throughput Computing (CHTC) at the University of Wisconsin{\textendash}Madison,
Open Science Grid (OSG),
Extreme Science and Engineering Discovery Environment (XSEDE),
Frontera computing project at the Texas Advanced Computing Center,
U.S. Department of Energy-National Energy Research Scientific Computing Center,
Particle astrophysics research computing center at the University of Maryland,
Institute for Cyber-Enabled Research at Michigan State University,
and Astroparticle physics computational facility at Marquette University;
Belgium {\textendash} Funds for Scientific Research (FRS-FNRS and FWO),
FWO Odysseus and Big Science programmes,
and Belgian Federal Science Policy Office (Belspo);
Germany {\textendash} Bundesministerium f{\"u}r Bildung und Forschung (BMBF),
Deutsche Forschungsgemeinschaft (DFG),
Helmholtz Alliance for Astroparticle Physics (HAP),
Initiative and Networking Fund of the Helmholtz Association,
Deutsches Elektronen Synchrotron (DESY),
and High Performance Computing cluster of the RWTH Aachen;
Sweden {\textendash} Swedish Research Council,
Swedish Polar Research Secretariat,
Swedish National Infrastructure for Computing (SNIC),
and Knut and Alice Wallenberg Foundation;
Australia {\textendash} Australian Research Council;
Canada {\textendash} Natural Sciences and Engineering Research Council of Canada,
Calcul Qu{\'e}bec, Compute Ontario, Canada Foundation for Innovation, WestGrid, and Compute Canada;
Denmark {\textendash} Villum Fonden and Carlsberg Foundation;
New Zealand {\textendash} Marsden Fund;
Japan {\textendash} Japan Society for Promotion of Science (JSPS)
and Institute for Global Prominent Research (IGPR) of Chiba University;
Korea {\textendash} National Research Foundation of Korea (NRF);
Switzerland {\textendash} Swiss National Science Foundation (SNSF);
United Kingdom {\textendash} Department of Physics, University of Oxford.
We also thank Ke Fang and Kohta Murase for useful comments.

\bibliography{main}{}
\bibliographystyle{aasjournal}

\end{document}